\documentstyle[11pt,newpasp,twoside,epsf]{article}
\def\edcomment#1{\iffalse\marginpar{\raggedright\sl#1\/}\else\relax\fi}
\reversemarginpar

\begin{document}

\title{Virial theorem analysis of 3D numerical simulations of
MHD self-gravitating turbulence} 
 \author{Mohsen Shadmehri}
\affil{Department of Physics, School of Science, Ferdowsi University,
Mashhad, Iran} 
\author{Enrique V\'azquez-Semadeni and Javier Ballesteros-Paredes}
\affil{Instituto de Astronom\' \i a, UNAM, Apdo. Postal 72-3(Xangari),
Morelia, Michoac\'an 58089, M\'exico} 

\begin{abstract}
We discuss the virial balance of all members of a cloud ensemble in
numerical simulations of self-gravitating MHD turbulence. We first
discuss the choice of 
reference frame for evaluating the terms entering the virial theorem (VT),
concluding that the balance of each cloud should be measured in its own
reference frame. We then report preliminary results suggesting that a)
the clouds are far from virial {\it equilibrium}, with the ``geometric'' 
(time derivative) terms dominating the VT. b) The surface terms in the
VT are as important as the volume ones, and tend to decrease the action
of the latter. c) This implies that gravitational binding should be
considered including the surface terms in the overall balance.
\end{abstract}

\section{Introduction}
The virial theorem (VT) (Chandrasekhar \&
Fermi 1953) provides a direct way of analyzing the
energy balance of a bounded region in a flow, describing the
effect of various forces either in driving changes in the structure of a
dynamical system or in determining the character of its equilibrium.

The virial theorem can be cast in either Eulerian or Lagrangian
form. The latter applies to a fluid parcel following the flow, i.e., the
volume $V$ and its bounding surface $S$ will generally be
time-dependent. The Eulerian version of the virial theorem (EVT) applies
to a fixed volume $V$ rather than a fixed mass (e.g., McKee \& Zweibel
1992). This is best suited for application to fixed-grid, Eulerian
numerical simulations. However, when considering a large region of the
ISM, clouds constitute an ensemble in which each one is {\it moving} and
{\it morphing} (see V\'azquez-Semadeni, this volume). Thus, for studying
the EVT we have two options: 

\noindent {(a) We can consider all clouds in the simulation, or ``lab'' frame
(Ballesteros-Paredes \& Vazquez-Semadeni 1997). However, in this
case, the clouds' bulk motions will appear as internal kinetic energy,
and the energy budget of the volume defining the cloud is not exactly
the energy budget of the cloud;} 

\noindent {(b) We can consider a different frame for each cloud, so
that the frame is moving with the cloud's center of 
mass velocity with respect to the lab frame, and has its
origin at the cloud's center of mass.}

In the present work, we briefly discuss the choice of reference frame,
and then present 
preliminary statistical results of virial balance in two three-dimensional
simulations of MHD isothermal turbulence with self-gravity at a
resolution of $100^3$ grid points, and with rms Mach number 2.2. One
simulation has a box size $L$ equal to the Jeans length $L_J$, and the
other has $L=2 L_J$.

\section{Virial Theorem in the lab and cloud frames}

The transition from the lab frame is not completely straightforward. We
first considered the possibility of staying in the lab frame, but taking 
a volume that moves with the velocity of the cloud's center of
mass, although maintaining a fixed shape in time. However, in this case,
the form of the EVT is altered, and several extra terms appear that have no
simple interpretation. Due to these difficulties, we finally have chosen to
completely move to each cloud's frame, as described above. In this case,
the EVT remains in its usual form, at the expense that plots for the
cloud ensemble comparing the various terms in the EVT contain data from
many different frames -- one for each cloud. The EVT is then
\begin{equation}
\frac{1}{2}\frac{d^{2}I_{E}}{dt^{2}}=2(\epsilon_{th}+\epsilon_{kin}-
\tau_{th}-\tau_{kin})+\epsilon_{mag}+\tau_{mag}+w-\frac{1}{2}\frac{d\Phi} 
{dt}, 
\end{equation}
where $I_{E}=\int_{V}\rho r^{2}dV$ is the moment of the inertia of the
cloud, $\epsilon_{th}=3/2 \int_{V} p dV$ is the thermal energy;
$\epsilon_{kin}= 1/2 \int_{V} \rho u^{2} dV$ is the kinetic energy,
$\tau_{th}=1/2 \oint_{S} x_{i}p n_{i} dS$ is the surface pressure
term, $\tau_{kin}=1/2 \oint_{S} x_{i}\rho u_{i}u_{j} n_{j} dS$ is
the surface kinetic term; $\epsilon_{mag}=1/8\pi \int_{V} B^{2} dV$ is
the magnetic energy and $\tau_{mag}=\oint_{S} x_{i}T_{ij} n_{j}
dS$ is the surface magnetic term, with $T_{ij}$ being the Maxwell stress
tensor;
$w=\int_{V} \rho x_{i} g_{i} dV$ is the gravitational term ({\it not}
equal to the gravitational energy), and
$\Phi=\oint_{S} \rho r^{2} u_{i} n_{i} dS$ is the flux of moment
of inertia through the surface of the cloud (sums over repeated indices
are assumed).

We define the clouds in the numerical simulations as connected sets of
pixels with densities larger than an arbitrary threshold value. In order
to measure only the contribution associated to the fluctuations, we
substract from the velocity the bulk, mass-averaged velocity of the cloud,
defined as ${\bf V}_{cm}=1/M \int_{V} \rho {\bf u} dV$, where $M$ is mass
of the cloud, and substract from the position of each pixel the position 
of the cloud's center of mass. We consider several density thresholds to 
enlarge the cloud ensemble, and consider that a cloud maintains its
identity as long as it does not split into several components upon
increasing the threshold.

\section{Preliminary Statistics}

We find the following preliminary results, consistent with those of
Ballesteros-Paredes \& V\'azquez-Semadeni (1997): The time derivative terms
(i.e., the LHS and the last term in the RHS of equation [1]) are
dominant in the overall virial balance, being much larger than the
remaining volume and surface terms (fig.1). We refer to these terms as
``geometrical'', since they correspond to the (time derivatives of the) mass
distribution in the cloud and its flux through the 
cloud's boundary.  This implies that, far from
being in quasi-hydrostatic equilibrium, the clouds are continually
changing shape and ``morphing''. {\bf 2.} The surface terms, which
are often neglected in virial-balance studies, have magnitudes comparable
to those of the volumetric ones (fig.2). {\bf 3.} However, the
surface and the volume terms do not cancel out exactly, leaving a
net contribution for shaping the clouds and balancing gravity. In
fact, we propose that the correct diagnostic for whether a cloud will
undergo gravitational collapse is
$|w|>2(\epsilon_{th}+\epsilon_{kin}-\tau_{th}-\tau_{kin})+\epsilon_{mag}+ 
\tau_{mag}$; otherwise, the cloud is transient. Indeed, in the simulation with $L=L_J$, two clouds 
are observed to be collapsing by the final time. These two clouds satisfy the above criterion (fig.3, left). 
Note that comparing the absolute value of gravitational term vs. only the volume term would suggest that none of the 
clouds be collapsing (fig.3, right), contrary to what occurs in the simulation. Thus, we conclude that {\it the 
correct diagnostic for determining gravitational binding must include
the contribution of the surface terms, or else the gravitational term is 
underestimated.}

\acknowledgements
This research has received partial financial support from Ferdowsi
University (to M.S.), and from CONACYT grant 27752-E to E.V.-S.

\begin{figure}
\plotone{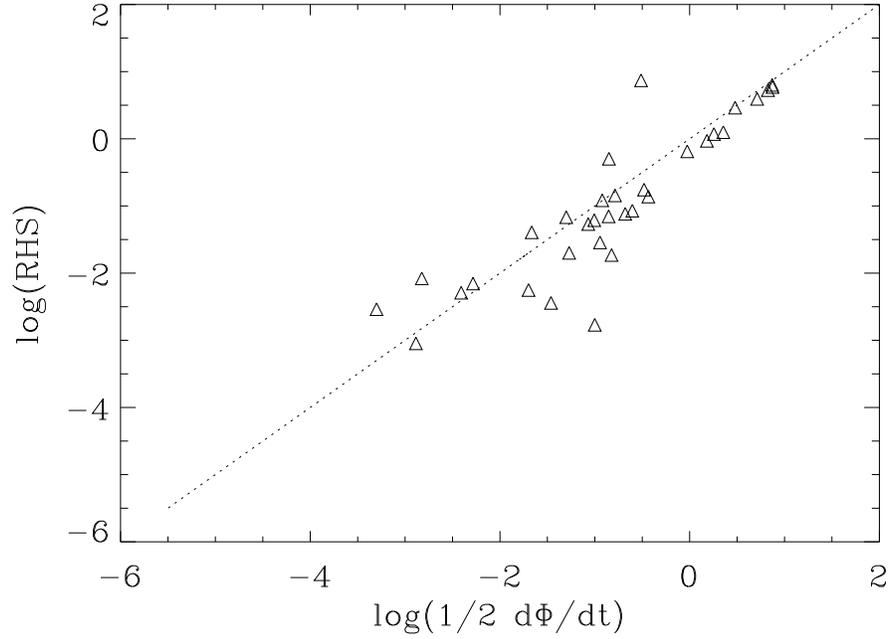}
\caption{The absolute value of RHS vs.\ the absolute value of
the $1/2d\Phi/dt$ of eq.\ (1). Their near equality shows that the time-derivative
term is the dominant one in the balance for the run with $L=L_J$.}
\end{figure}

\begin{figure}
\plottwo{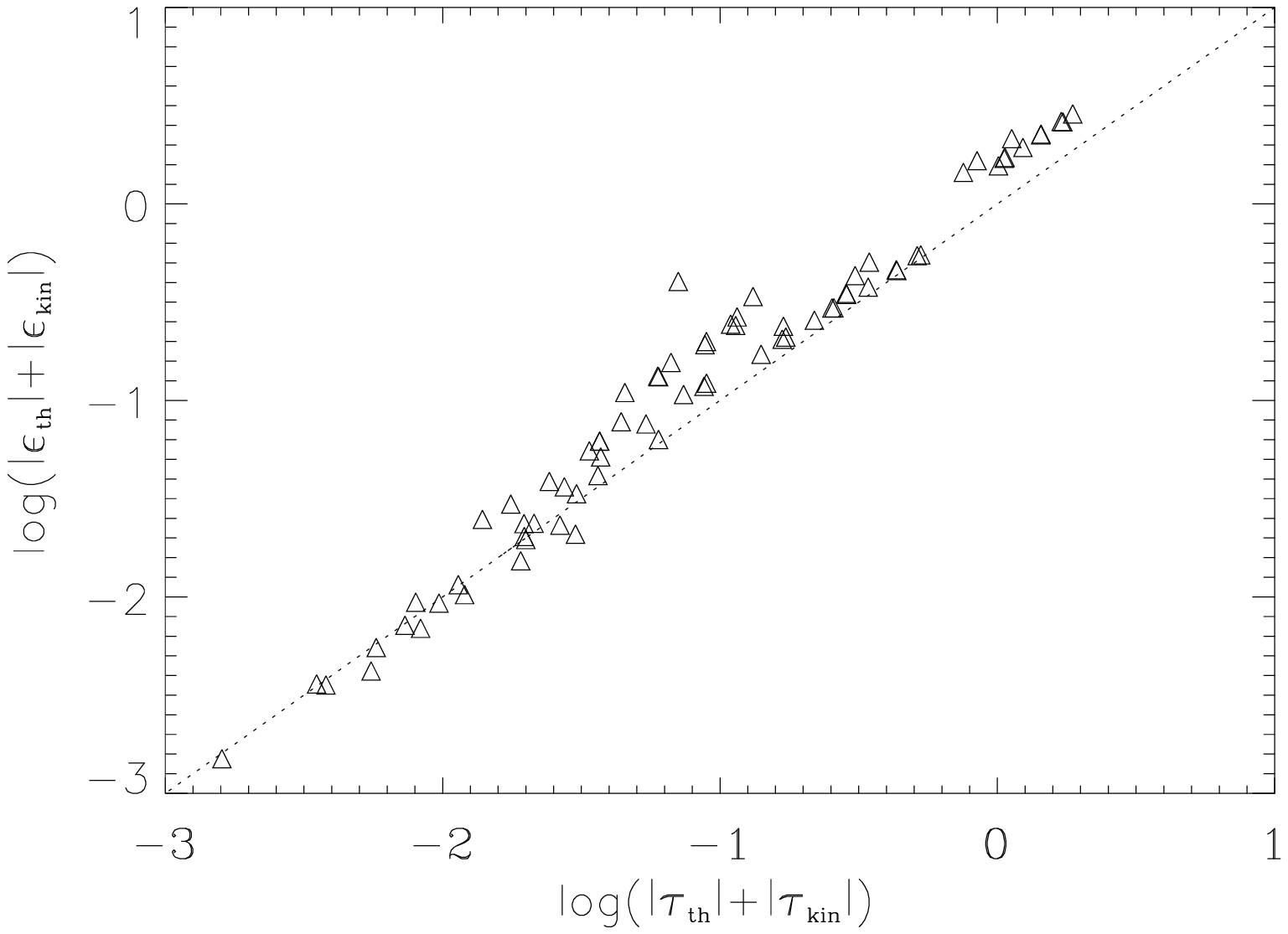}{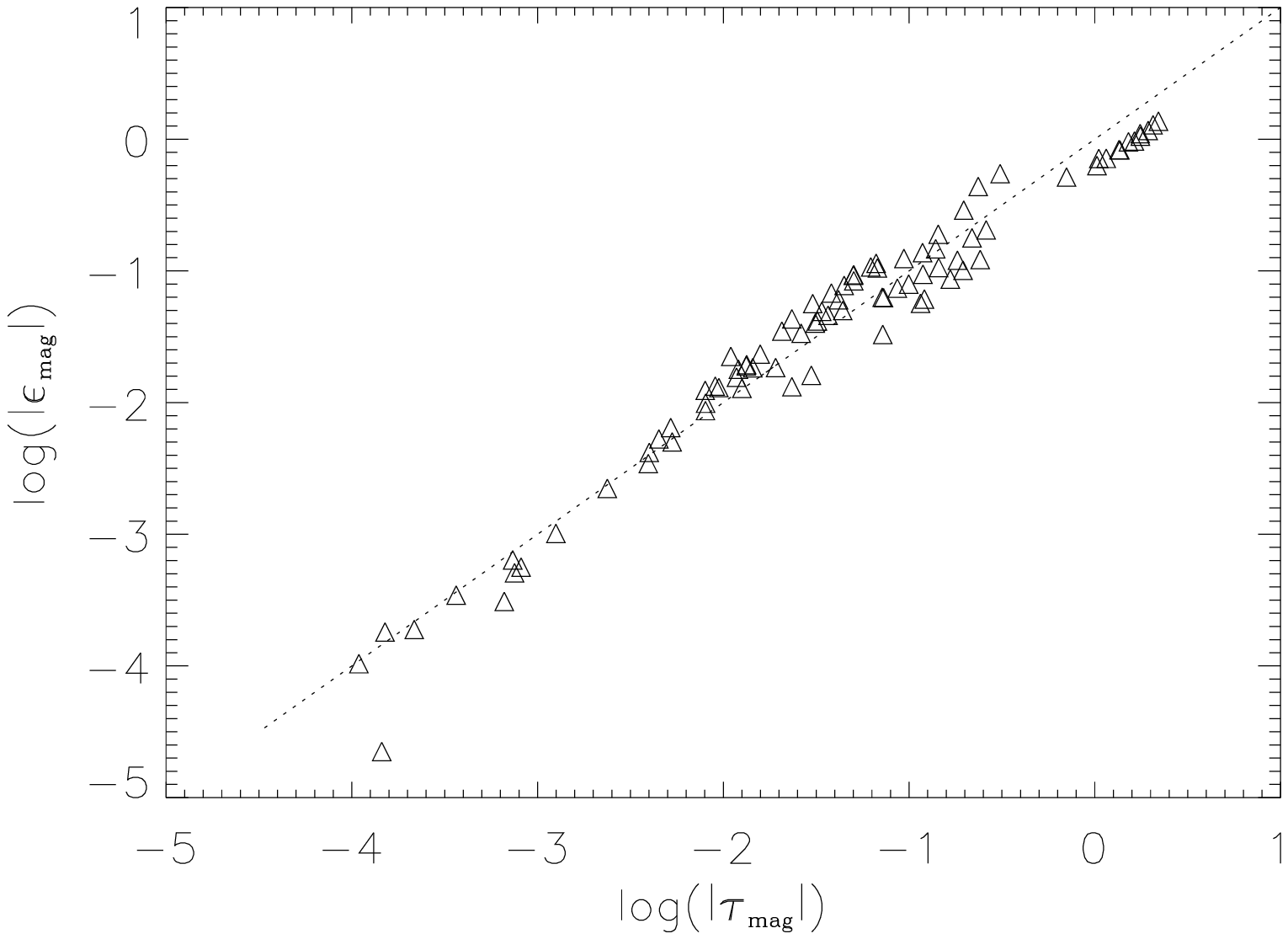}
\caption{Volume-vs.-surface balance of the thermal-plus-kinetic
terms (left) and the magnetic terms (right) for the run with $L=L_J$. The near equality of both kinds of terms indicates 
that the contribution of the surface terms is comparable to that of the
volume terms, implying partial 
cancellation.} 
\end{figure}

\begin{figure}
\plottwo{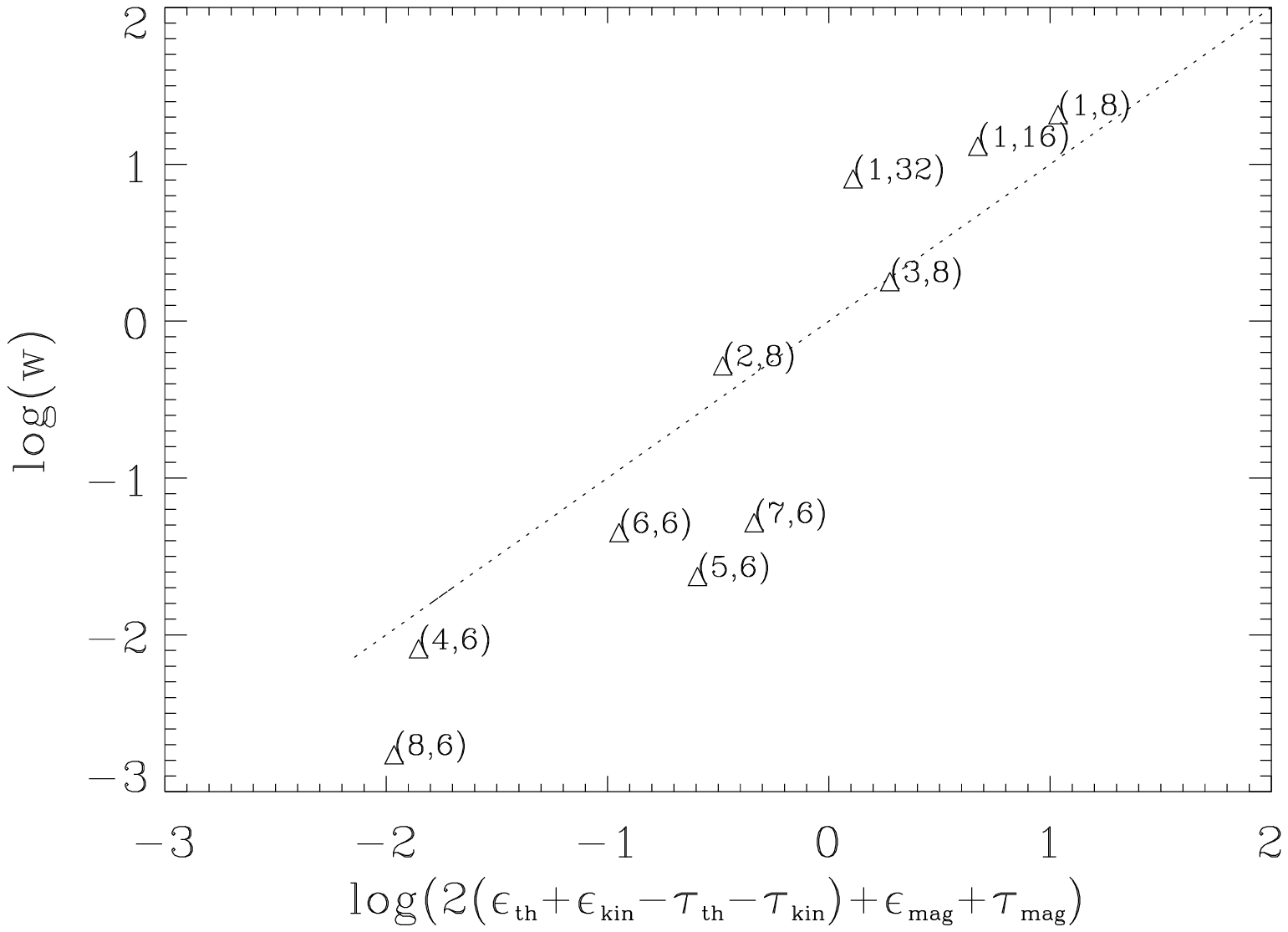}{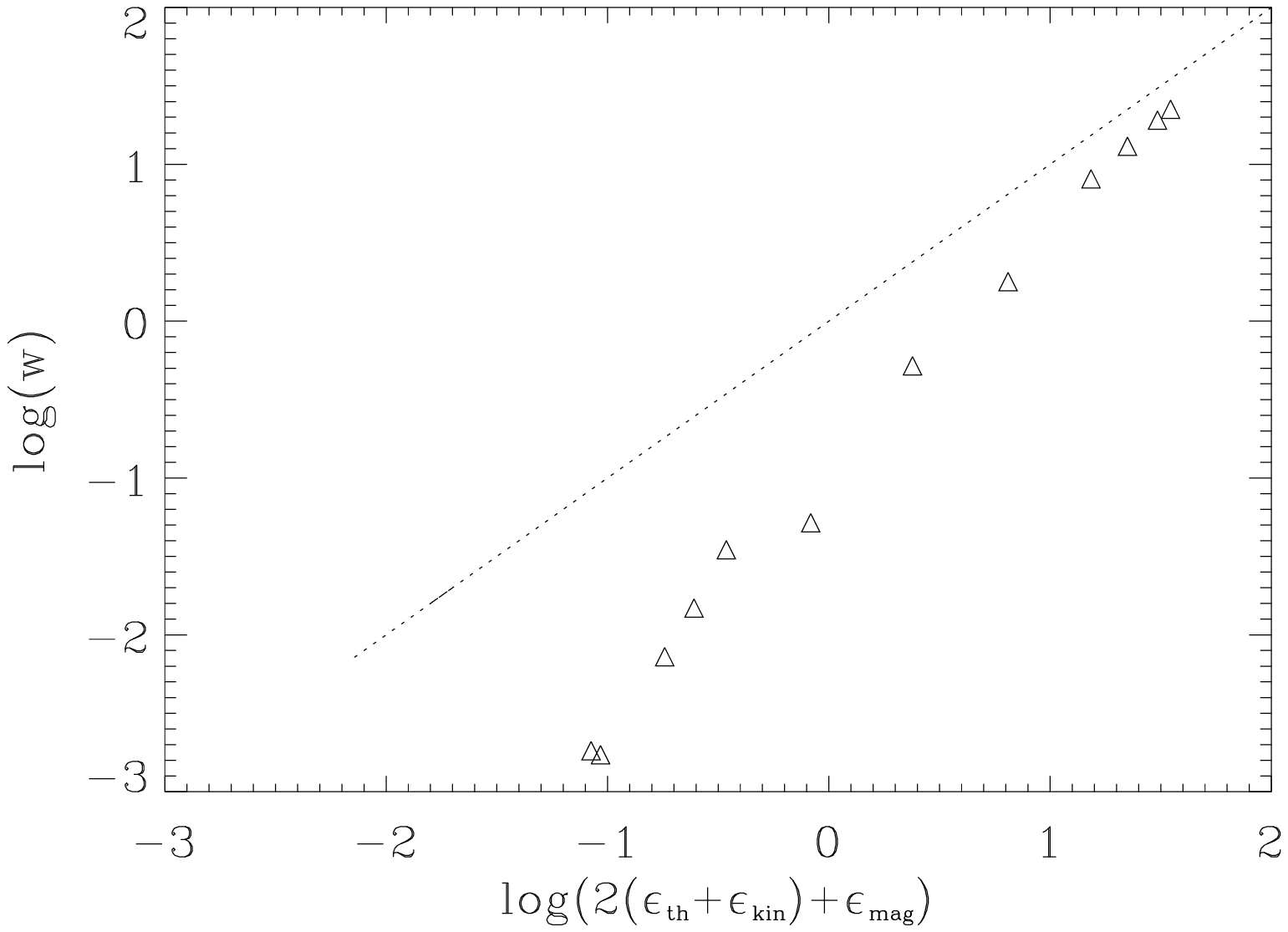}
\caption{The gravitational term $w$ vs.\ the RHS of eq.\ (1) without the 
$1/2 d \Phi/dt$ term for the simulation with $L=2L_J$. In the plot on the left, 
we label each cloud by a pair of numbers, one identifying the cloud, and the other giving the threshold density 
defining the cloud. This plot accurately represents the fact that only two clouds are 
collapsing in the simulation (i.e., clouds 1 and 2), while the other clouds are not collapsing. Comparing the absolute 
value of gravitational term vs. only the volume term would suggest that none of the clouds should be collapsing (right), contrary to what occurs 
in the simulation.}
\end{figure}

\end{document}